\documentclass{kluwer}    

\usepackage{epsfig}
\newdisplay{guess}{Conjecture}

\begin{document}                                                                                   
\begin{article}
\begin{opening}         
\title{XTE J1118+480: Clues on the Nature of the Accretion Flow from the Optical Variability 
} 
\author{A. \surname{Merloni}$^1$} 
\author{T. \surname{Di Matteo}$^2$\thanks{Chandra fellow}}
\author{A. C. \surname{Fabian}$^1$} 
\runningauthor{Merloni, Di Matteo and Fabian}
\runningtitle{XTE J1118+480: Optical Variability
 and the Nature of the Accretion Flow}
\institute{{\rm 1} Institute of Astronomy, Madingley Road, Cambridge, CB3 0HA, UK
\\ {\rm 2} Harvard-Smithsonian Center for Astrophysics, 60 Golden St., 
Cambridge, MA 02138, USA}

\begin{abstract}
We show how the simultaneous presence of a strong quasi periodic oscillation 
(QPO) of
period $\sim$10 seconds in the optical and X-ray lightcurves of the
X-ray transient XTE J1118+480 can be used to obtain information about the 
nature of the accretion flow around the source.
The unusually high optical-to-X-ray flux ratio and the QPO 
observed simultaneously in both energy bands 
suggest that a significant fraction of
the optical flux might originate close to the central source,
where most of the X-rays are produced and be indicative of a 
magnetically dominated corona.
We also show how the temporal evolution of the QPO can provide us with 
information on both the inner radius and the 
viscous properties of the optically thick accretion disc.

\end{abstract}

\end{opening}           

\section{Introduction: The Magnetic Corona Model}  

The transient X-ray source XTE J1118+480 was
discovered with the RXTE All-Sky Monitor on March
29, 2000 (Remillard et al. 2000). 
At the peak of its outburst, it showed an X-ray spectrum typical of black hole
candidates (BHC) in their {\it hard} state, with a prominent power-law
component (photon index of about 1.8).  
The 13th
magnitude optical counterpart, discovered by Uemura, Kato \& Yamaoka
(2000), exhibited a spectrum fairly typical of X-ray Novae in outburst
(Dubus et al. 2000 and references therein).
A strong QPO with frequency $\nu_{\rm QPO} \sim 0.1$ Hz
has been found
in the X-ray power density spectrum (PDS) of the source (Wood et al. 2000);  
more surprisingly, the optical lightcurves also showed a
prominent QPO at the same frequency 
(Haswell et al. 2000).
The QPO frequency drifted to higher values systematically in both bands
for about two months.

As discussed in detail in Merloni, Di Metteo \& Fabian (2000) (MDF), the 
high optical-to-X-ray flux ratio and the optical/UV variability 
argues for the presence of significant
cyclo-synchrotron (CS) radiation from active regions in a 
 magnetically dominated corona 
in this source.
Assuming that a fraction of the observed optical 
flux corresponding to its oscillating  part ($\sim 30\%$) is due to 
CS emission 
from a magnetized corona, we modeled the spectrum 
taking into account 
reprocessing of coronal radiation in the accretion disc (according 
to the height of the active regions) and all 
the relevant radiative processes (see MDF for details).
 The disc was assumed to extend down to the
innermost stable orbit of a non-rotating black hole ($R_{\rm
in}=6GM/c^2=3R_{\rm S}$).
We found that, 
in order to match the high (compared to X-rays) optical flux, 
both a low accretion rate 
($\dot M/\dot M_{\rm Edd}=\dot m\simeq 0.01$), 
and a high fraction of accretion power 
dissipated in the corona ($f>0.9$) are needed; 
also the active regions need to be 
high above the disc (at about $10 R_{\rm S}$)\footnote{
Alternatively, a good agreement with the data could be 
achieved if the active  
regions move relativistically away from the disc 
(MDF, see also 
Markoff, Falcke \& Fender 2000 for a jet model of the SED)}.

Hynes et al. (2000) reporting on {\it EUVE} observations of the source
showed that the low measured EUV flux is
inconsistent with an optically thick disc extending down to radii
$R < 1000 R_{\rm S}$ 
(but see Dubus et al. 2000 for a different estimate of the 
EUV flux). Based on this, they favor an inner advection-dominated flow (ADAF)
surrounded by a geometrically 
thin disc truncated at hundreds of $R_{\rm S}$.
In the next section we show how to use the variability information to put constraints on the inner disc location.

\section{The QPO Viscous Evolution}

Beside the peaked QPO around 0.1 Hz, the observed PDS of XTE J1118+480 
exhibits a broader feature around 1 Hz; plotted one
against the other, these frequencies fall remarkably well on the correlation
among QPOs discovered by Psaltis, Belloni \& van der Klis (1999) 
in the PDS of BHC and neutron stars.
Such a correlation is likely to reflect a general property of the 
accretion flows around these objects. Here we assume that the QPO observed
in XTE J1118+480 is 
produced at a specific distance ($R_{\rm t}=r_{\rm t} R_{\rm S}$) 
where a discontinuity of the flow occurs, 
as in Psaltis \& Norman (2000). The QPO frequency must be 
essentially determined by the relativistic dynamical frequencies of the
system (Psaltis \& Norman 2000; Stella, Vietri \& Morsink 1999). 

\begin{figure}
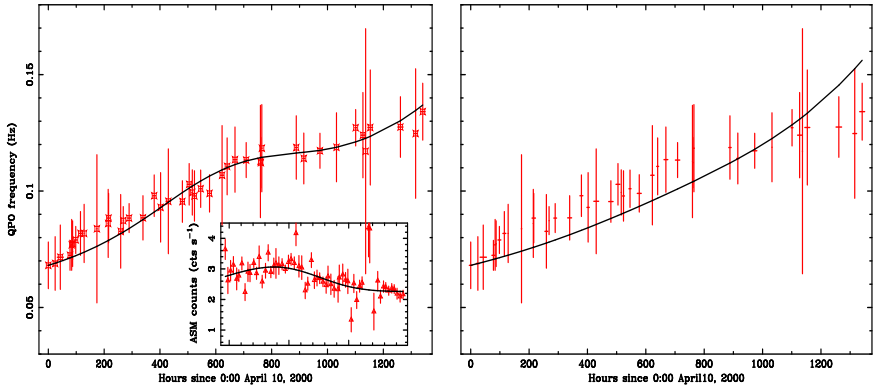

\tabcapfont
\centerline{
\begin{tabular}{c|@{\hspace {0.7pc}}c}
(a) $\nu_{\rm QPO}$=$2\nu_{\rm LT}$; rad. p.; $\chi^2/{\rm dof}$=$0.21$; &
(b) $\nu_{\rm QPO}$=$\nu_{\rm Kep}$; gas p.; $\chi^2/{\rm dof}$=$0.80$;  \\
$\alpha [\dot m (1-f)]^2 \simeq 4.5\times 10^{-8} a^{7/6} m^{-1/6}$
&
$\alpha^{4/5} [\dot m^2 (1-f)]^{1/5} \simeq 2.5\times 10^{-4} m^{4/15}$
\\ 
\end{tabular}
}
\centerline{
\begin{tabular}{c@{\hspace {3pc}}c}
\includegraphics[width=11pc]{fig1gr.eps} &
\includegraphics[width=11pc]{kg.eps} \\
\end{tabular}
}
\caption{Fit to the observed QPO evolution (RXTE and USA data, Wood et al 2000)
 for (a) a QPO
produced by relativistic LT precession in a radiation pressure dominated disc
and (b) a QPO produced by modulation with Keplerian frequency 
in a gas pressure dominated disc. 
The error bars are given by the QPO FWHM, which is likely to 
reflect the uncertainties in the location where the oscillation are 
excited. In the inset the ASM lightcurve of XTE J1118+480 (triangles): the 
continuous line is a SPLINE fit to its smooth variation 
(flares are excluded), which is assumed to reflect long term variations of 
the total accretion rate.   
}
\end{figure}

XTE J1118+480 observations offer the opportunity to track the QPO
evolution over a long period of time (Wood et al. 2000).
To model such evolution, we calculate $R_{\rm t}(t)$ integrating
the equation of the viscous 
evolution of a standard Shakura-Sunyaev disc: 
\begin{equation}
\frac{\partial R_{\rm t}}{\partial t}=-\frac{3}{2} \alpha \left(\frac{H}{R_{\rm t}}\right)^2 \left(\frac{GM}{R_{\rm t}^3}\right)^{1/2}J^{-1},
\end{equation}
where $J$=$1-\sqrt{3/r_{\rm t}}$, $\alpha$ is the viscous 
parameter and the 
disc scaleheight $H/R_{\rm t}\simeq 9 \dot m J (1-f) r_{\rm t}^{-1}$ for a 
radiation pressure dominated disc and 
$H/R_{\rm t}\simeq 0.02 (\alpha m)^{-1/10} (\dot m J)^{-1/5} (1-f)^{-1/10} r_{\rm t}^{1/20}$
for a gas pressure dominated one (Merloni, Fabian \& Ross 2000).
We test two different hypothesis on the origin of the 
QPO (and the accretion flow geometry):

(a) According to the relativistic precession interpretation of the observed 
correlation (obeyed by XTE J1118+480) 
$\nu_{\rm QPO}$ is twice the nodal precession frequency:
$\nu_{\rm QPO}$=$2\nu_{\rm nod} \simeq 1.62 \times 10^4 (a/m) r_{\rm t}^{-3}$ Hz, implying $r_{\rm t} \sim 50 (a/m)^{1/3}$ ($a$ is the BH 
 angular momentum, $m$ its mass in Solar mass units). In this case the 
radiation pressure solution, appropriate for the inner 
part of an optically thick disc, should be used in Eq (1).

(b) Alternatively, the observed low frequency QPO can be due to
a Keplerian modulation of the flow: 
$\nu_{\rm QPO}$=$\nu_{\rm Kep} \simeq 1.14 \times 10^4 m^{-1} r_{\rm t}^{-3/2}$ Hz, so that
$r_{\rm t} \sim 2000 \, m^{-2/3}$, 
and the gas pressure solution must be used in this case.

Furthermore, we assume that the smooth variations of the 
X-ray ASM luminosity reflects a modulation of the total 
accretion rate so that $\dot m(t)$=$g(t)\dot m_0 $, where $g(t)$ 
is the continuous function shown in the inset in Fig. 1.
We integrate Eq. (1) and 
fitted the temporal evolution of the QPO frequency with the resulting
$\nu_{\rm QPO}(R_{\rm t}(t))$   
for the two cases and show the results 
in Fig. 1. Both fits are acceptable, but the 
relativistic precession model (with $r_{\rm t}$=$O(10)$) 
explains the QPO evolution better than the Keplerian one  
(with $r_{\rm t}$=$O(1000)$).
Finally, from such fits we can obtain an estimate of the viscous
parameters of the optically thick accretion disc 
($\alpha$, $\dot m$, $f$, through the 
products shown in Fig. 1) independent on any
spectral analysis. The estimate from MDF ($\dot m f\simeq 10^{-3}$) 
based on the magnetic corona model is also consistent with 
case (a) (with $\alpha \sim 0.01$), while for case (b) it would 
imply an unreasonably low viscosity.  

\section{Conclusions}
 
The main rationale for our model is the simultaneous presence of a
strong QPO in the optical and X-ray lightcurves of XTE J1118+480, 
suggesting that the fluxes in the two bands both originate from the same
region in the inner part of the accretion flow. Self-absorbed
CS emission is the natural candidate to explain the
optical variability.
Such emission is expected in any magnetic corona model, and the
inferred magnetic field value ($B\approx 2 \times 10^6$ G) is the one
predicted to arise when the source is in the {\it hard} state (MDF). 
We have also shown how the evolution of the QPO 
can be satisfactory accounted for if the QPO is 
produced  by a 
modulation  
whose frequency is equal to twice the relativistic 
nodal precession in a radiation pressure dominated disc and the location 
of such modulation (that is perhaps where
the inner accretion disc undergoes a transition and is  
{\it at most} a few tens of 
Schwarzschild radii) viscously drift inwards.


\end{article}
\end{document}